\documentclass[manuscript]{aastex6}

\usepackage{breakurl}
\usepackage{amsmath}
\usepackage{cases}

\shorttitle{EVOLUTION OF ALFV\'ENIC FLUCTUATIONS INSIDE AN ICME}
\shortauthors{LI ET AL.}

\begin{document}


\title{Evolution of Alfv\'enic fluctuations inside an interplanetary coronal mass ejection and their contributions to local plasma heating: Joint observations from 1.0 AU to 5.4 AU}

\author{Hui Li \altaffilmark{1}, Chi Wang \altaffilmark{1, 4}, John D. Richardson \altaffilmark{2} and Cui Tu \altaffilmark{3}}


\altaffiltext{1}{State Key Laboratory of Space Weather, National Space Science Center, CAS, Beijing, 100190, China; \url{hli@nssc.ac.cn}}

\altaffiltext{2}{Kavli Institute for Astrophysics and Space Research, Massachusetts Institute of Technology, Cambridge, MA, USA}

\altaffiltext{3}{Laboratory of Near Space Environment, National Space Science Center, CAS, Beijing, 100190, China}

\altaffiltext{4}{University of Chinese Academy of Sciences, Beijing, 100049, China.}



\begin{abstract}

Directly tracking an interplanetary coronal mass ejection (ICME) by widely separated spacecrafts is a great challenge. However, such an event could provide us a good opportunity to study the evolution of embedded Alfv\'enic fluctuations (AFs) inside ICME and their contributions to local plasma heating directly. In this study, an ICME observed by \textit{Wind} at 1.0 au on March 4-6 1998 is tracked to the location of \textit{Ulysess} at 5.4 au. AFs are commonly found inside the ICME at 1.0 au, with an occurrence rate of 21.7\% and at broadband frequencies from 4$\times 10^{-4}$ to 5$\times 10^{-2}$ Hz. When the ICME propagates to 5.4 au, the Aflv\'enicity decreases significantly, and AFs are rare and only found at few localized frequencies with the occurrence rate decreasing to 3.0\%. At the same time, the magnetic field intensity at the AF-rich region has an extra magnetic dissipation except ICME expansion effect. The energetics of the ICME at different radial distance is also investigated here. Under similar magnetic field intensity situations at 1.0 au, the turbulence cascade rate at the AF-rich region is much larger than the value at the AF-lack region. Moreover, it can maintain as the decrease of magnetic field intensity if there is lack of AFs. However, when there exists many AFs, it reduces significantly as the AFs disappear. The turbulence cascade dissipation rate within the ICME is inferred to be 1622.3 $J\cdot kg^{-1}\cdot s^{-1}$, which satisfies the requirement of local ICME plasma heating rate, 1653.2 $J\cdot kg^{-1}\cdot s^{-1}$. We suggest that AF dissipation is responsible for extra magnetic dissipation and local plasma heating inside ICME.

\end{abstract}


\section{Introduction}
\label{sec:intro}

Coronal mass ejections (CMEs) are spectacular eruptions in the solar atmosphere \citep[e.g.][]{Kunow et al 2006,Gopalswamy 2010}. The interplanetary manifestations or the heliospheric counterparts of CMEs are referred to as Interplanetary coronal mass ejections (ICMEs) \citep[e.g.][]{Gosling 1990}, which are a key link between activities at the Sun and disturbances in the heliosphere. It is well known that ICMEs are important drivers of interplanetary shocks and disastrous space weather events, such as geomagnetic storms \citep[e.g.][and the references therein]{Richardson and Cane 2011}. Generally, many low-frequency magnetohydrodynamics waves waves, such as Alfv\'en waves (AWs) or Alfv\'enic fluctuations (AFs), fast and slow mode of magneto-acoustic waves, could be generated due to magnetic reconnection or catastrophe processes during the CME initiation \citep{Kopp and Pneuman 1976,Antiochos et al 1999,Chen and Shibata 2000}. In addition, some plasma waves could also be generated by the interactions between ICME and the ambient solar wind.

Compared to the massive studies on the macro-structures of ICMEs, the micro-states of ICMEs, especially the wave-particle behaviors which may contribute to particle energization and/or kinetic processes, have not been well discussed. To our best knowledge, only a few works were carried out to investigate the wave phenomena inside ICMEs. Some intense high frequency waves were found inside ICMEs, such as possible ion acoustic waves \citep{Fainberg et al 1996,Lin et al 1999,Thejappa and MacDowall 2001}, whistler waves and Langmuir waves \citep{Moullard et al 2001}. \citet{Siu-Tapia et al 2015} recent found low frequency waves inside 8 isolated magnetic clouds based on the \textit{STEREO} observations. \citet{Zhao et al 2017} studied 7807 electromagnetic cyclotron waves (ECWs) near the proton cyclotron frequency in and around 120 magnetic clouds during 2007--2013. For ultra-low frequency waves, well less than the proton cyclotron frequency, some authors also discussed AWs or AFs inside ICMEs. \citet{Marsch et al 2009} and \citet{Yao et al 2010} found possible AF events inside two ICMEs detected by \textit{Helios 2} at 0.7 au and 0.3 au, respectively. \citet{Liang et al 2012} later reported a clear AF event with 1-hr duration inside an ICME at 1 AU. Expect case studies, two statistical surveys have been performed to our knowledge. \citet{Li et al 2013} investigated 27 ICMEs near 1 au, finding that AWs exist continuously for 8 ICMEs, fast mode waves exist in the sheath of 13 ICMEs, and slow mode waves exist in all events. \citet{Li et al 2016a} extended the statistical study out to 6 au based on the 33 ICMEs observed by \textit{Voyager 2}. They confirmed the existence of AFs inside ICMEs, and concluded that the percentage of AF duration decays linearly in general as ICMEs expand and move outward. 

The evolution of ICMEs in the heliosphere is regarded as one of the fundamental issues in space physics. It is of great significance to study the properties and evolution characteristics of plasma waves inside ICMEs. Firstly, the spatial distributions of AFs inside ICMEs could give some clues of CME initiation processes and triggering mechanism \citep{Liang et al 2012}. Secondly, the evolution and dissipation of plasma waves inside ICMEs are helpful to understand the local plasma heating \citep{Tu and Marsch 1995,Kasper et al 2008,Wang et al 2014} during the non-adiabatic expansion of ICMEs between 0.3 $\sim$ 30 au \citep{Wang and Richardson 2004,Richardson et al 2006}. \citet{Liu et al 2006} have inferred that the nonlinear cascade of low-frequency AFs caused the magnetic dissipation within ICMEs, which is sufficient to explain the in-situ heating of ICME plasma. \citet{Li et al 2016a} later found similar ``W"-shaped distributions of AF occurrence and the proton temperature inside ICMEs, and confirmed the significant contribution of AFs on local ICME heating. 

Among plasma waves, AWs or AFs are of interest in the present study because they are the most common wave mode in the solar wind and inside ICMEs \citep{Bruno et al 2006,Li et al 2016a,Li et al 2016b}. Theoretical studies suggested AF dissipation contribute to ICME plasma heating. Statistical observation surveys provide some indirect evidences to support such statement. Directly tracking a specific ICME through the heliosphere by widely separated spacecrafts would provide us a good opportunity to study the evolution characteristics of embedded AFs and their contributions to ICME plasma heating, but it has never been done before. In this study, an ICME observed by \textit{Wind} at 1 au on March 4-6 1998 is tracked to the location of \textit{Ulysess} at 5.4 au. The goal is to give us more comprehensive understandings on this issue.

\section{Evolution of Alfv\'enic fluctuations inside the ICME}
\label{sec:obser}
We analyzed data sets for interplanetary magnetic field and solar wind plasma from \textit{Wind} and \textit{Ulysess} spacecraft. For \textit{Wind} spacecraft, the magnetic field data with a temporal cadence of 0.092 seconds are used from the Magnetic Field Investigation \citep[MFI; ][]{Lepping et al 1995}, and the solar wind plasma data with a temporal cadence of 3 seconds are used from the three-dimensional Plasma and Energetic Particle Investigation \citep[3DP; ][]{Lin et al 1995}. All the data from \textit{Wind} spacecraft are in the Geocentric Solar Ecliptic (GSE) coordinates. For \textit{Ulysess} spacecraft, the magnetic field data with a temporal resolution of 1 second are used from the Vector Helium Magnetometer \citep[VHM; ][]{Balogh et al 1992}, and the solar wind plasma data with a temporal resolution of 4 minutes are used from the Solar Wind Observations Over the Poles of the Sun \citep[SWOOPS; ][]{Bame et al 1992}. All the data from \textit{Ulysess} spacecraft are in the heliographic radial tangential normal (RTN) coordinate system.

\begin{figure}[htbp!]
\centering
\noindent\includegraphics[width=19pc]{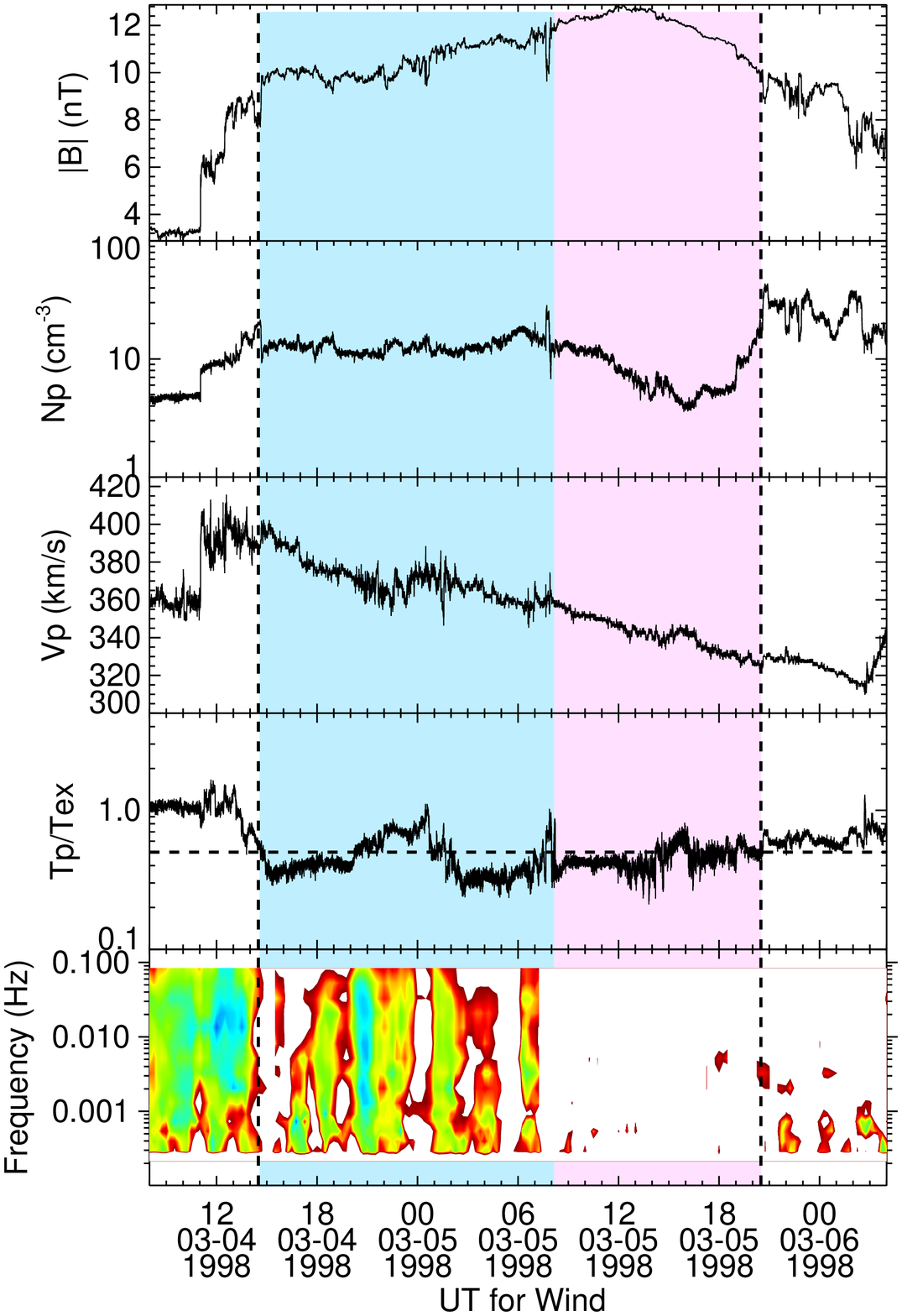}
\noindent\includegraphics[width=21.15pc]{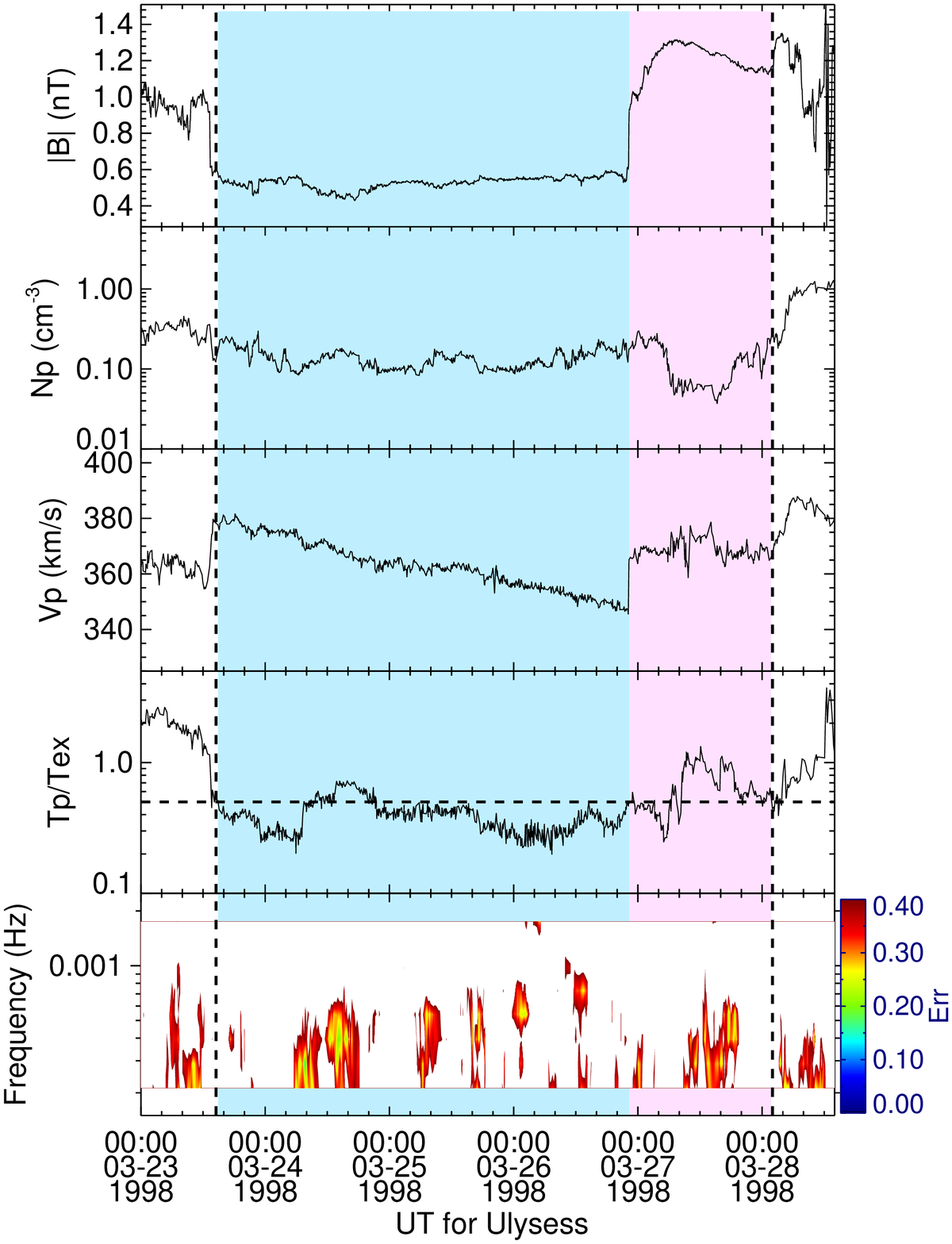}
\caption{Overview of the ICME (between two vertical dashed lines) observed by \textit{Wind} at 1 au and \textit{Ulysess} at 5.4 au. From top to bottom, the panels show the magnetic field strength ($|B|$), the proton number density ($N_p$), the solar wind bulk speed ($V_p$), the ratio of the observed to the expected temperature ($T_p/T_{ex}$), and the parameter representing the Alfv\'enicity ($E_{rr}$), respectively. The sky blue area represents the AF-rich region, while the light pink area denotes the AF-lack region.}
\label{ICME}
\end{figure}

The ICME was first observed by \textit{Wind} at 1 au on March 4-6 1998 and then passed through \textit{Ulysess} at 5.4 au on March 23-28 1998. During this period, these two spacecrafts lined up near the ecliptic plane with the latitudinal separation of $\sim$ 2$^\circ$, and longitude separation of $\sim$ 6$^\circ$. \citet{Du et al 2007} have confirmed these two ICMEs are the same one with the same solar origin by using a 1D MHD solar wind model and the Grad-Shafranov reconstruction technique. Such a great alignment between the Sun and both the spacecrafts provide us a unique opportunity to investigate the same ICME at two different evolution stages in the heliosphere. \citet{Skoug et al 2000}, \citet{Du et al 2007}, and \citet{Nakwacki et al 2011} have studied the dynamical evolution of the magnetic cloud macro-structure from the Sun to 5.4 au by analyzing the joint observations of this ICME event. 

Figure~\ref{ICME} shows the overview of the ICME observed by \textit{Wind} and \textit{Ulysess}. The two vertical dashed lines represent the start and end time of the ICME, which are consistent with the results of \citet{Nakwacki et al 2011}. The shock sheath is not included here. The primary criterion of an ICME is that the proton temperature ($T_p$) is lower than the expected temperature ($T_{ex}$) by a factor of 2. $T_{ex}$ is calculated from the relationship derived by \citet{Lopez 1987}. From \textit{Wind} observations, some other ICME signatures are clear, including: the enhancement of the magnetic field ($|B|$); the extreme increase of the proton number density ($N_p$); the monotonic declining of solar wind bulk speed ($V_p$). The parameter, $E_{rr}$ is introduced by \citet{Li et al 2016c} to represent the goodness of the degree of the Alfv\'enicity. The AFs are defined as the intervals with $E_{rr} \le $ 0.3 in this work. The time-frequency distribution of $E_{rr}$ reveals that there exists many relatively pure AFs at broadband frequencies inside the ICME at 1 au, from 4$\times 10^{-4}$ to 5$\times 10^{-2}$ Hz. According to the occurrence of AFs, the ICME could be divided into two regions. The first one contains many AFs with high degree of Alfv\'enicity at broadband frequencies, which is referred as the AF-rich region and represented in sky blue. The other part is lack of AFs which is thus called the AF-lack region and represented in pink. In general, the ICME has a duration of 30 hours with the width of 0.26 au. The average $V_p$, $|B|$, $N_p$, and $T_p$ of the ICME at 1 au is 358.9 km/s, 11.1 nT, 11.12 $cm^{-3}$, and 23802 K, respectively. 

When the ICME propagates to \textit{Ulysess}, the typical ICME signatures at 1.0 au are blurred due to some interactions with the ambient solar wind. The AFs are only found at very localized frequencies. Compared to the ICME observed by \textit{Wind} spacecraft, it has a longer duration of 107.5 hours, with the width of 0.94 au. The ICME speed has a slight increase to 364.4 km/s, while the magnetic field intensity, the number density, and the proton temperature decreases to 0.7 nT, 0.135 $cm^{-3}$, and 5370 K, respectively. Meanwhile, the AFs are only found at very localized frequencies, especially for the previous AF-rich region at 1 au. In addition, the magnetic field intensity at the previous AF-rich region is much less than that at the previous AF-lack region.

\begin{figure}[htbp!]
\centering
\noindent\includegraphics[width=18pc,angle=90]{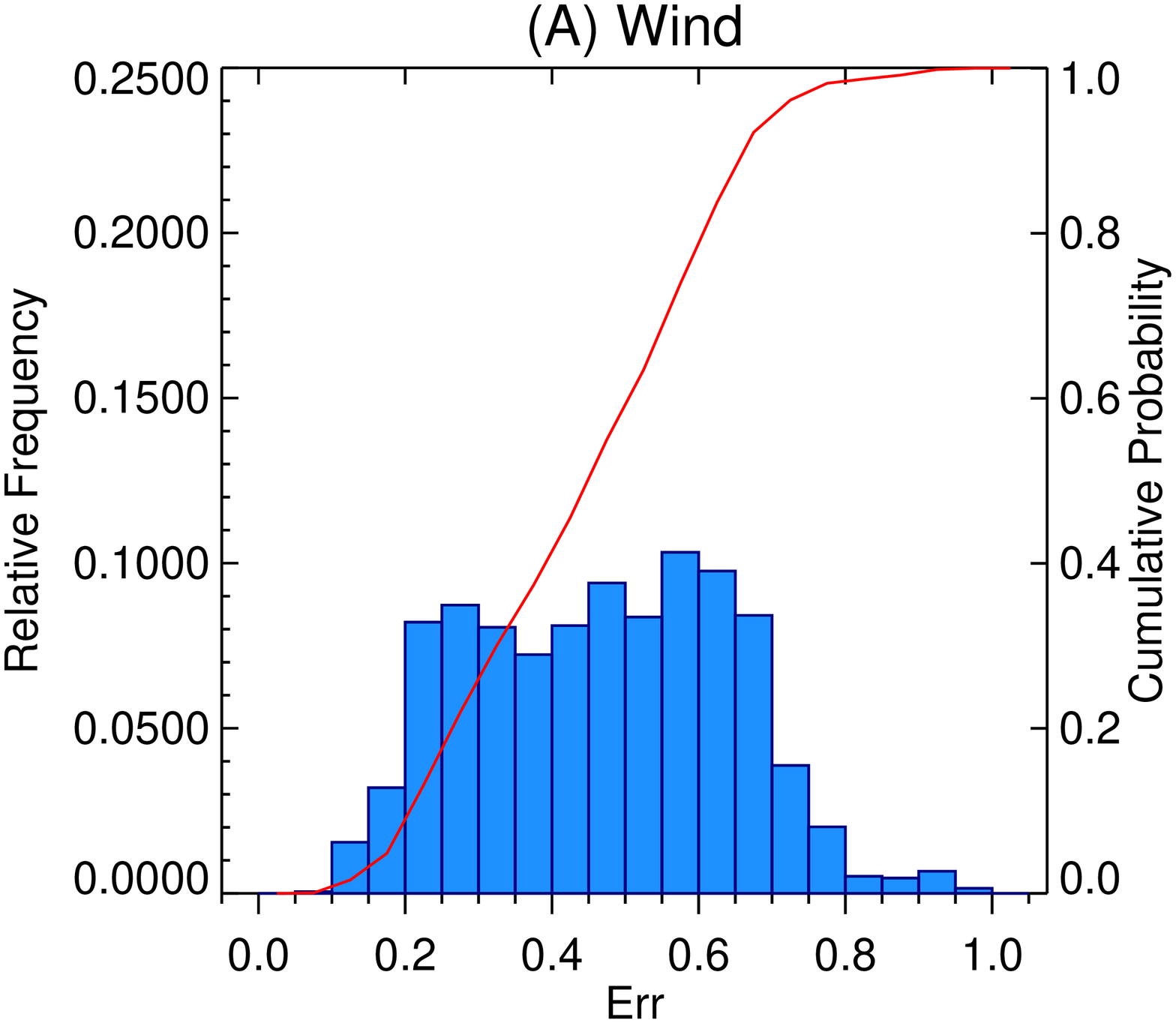}
\noindent\includegraphics[width=18.5pc,angle=0]{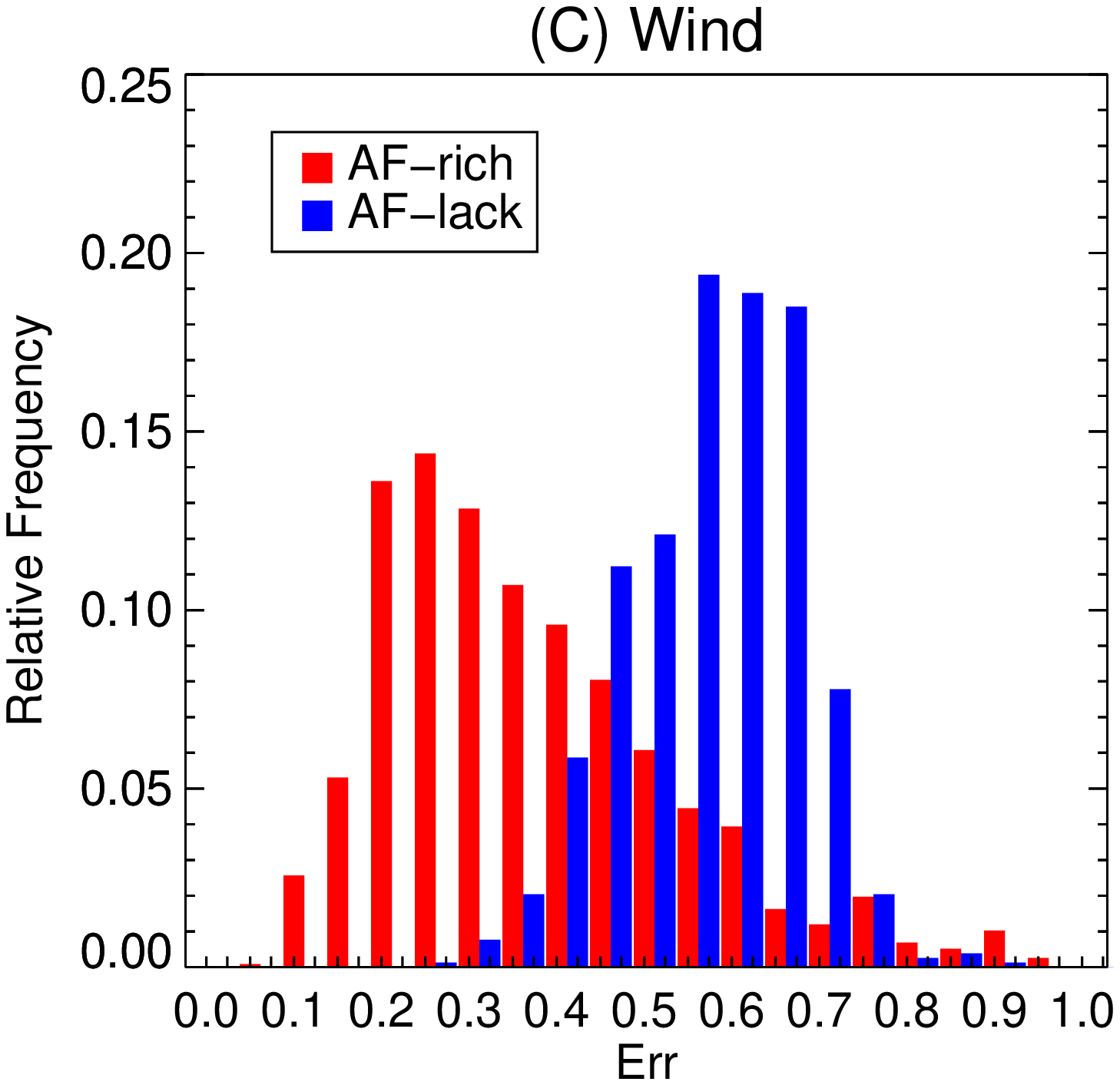}
\noindent\includegraphics[width=18pc,angle=90]{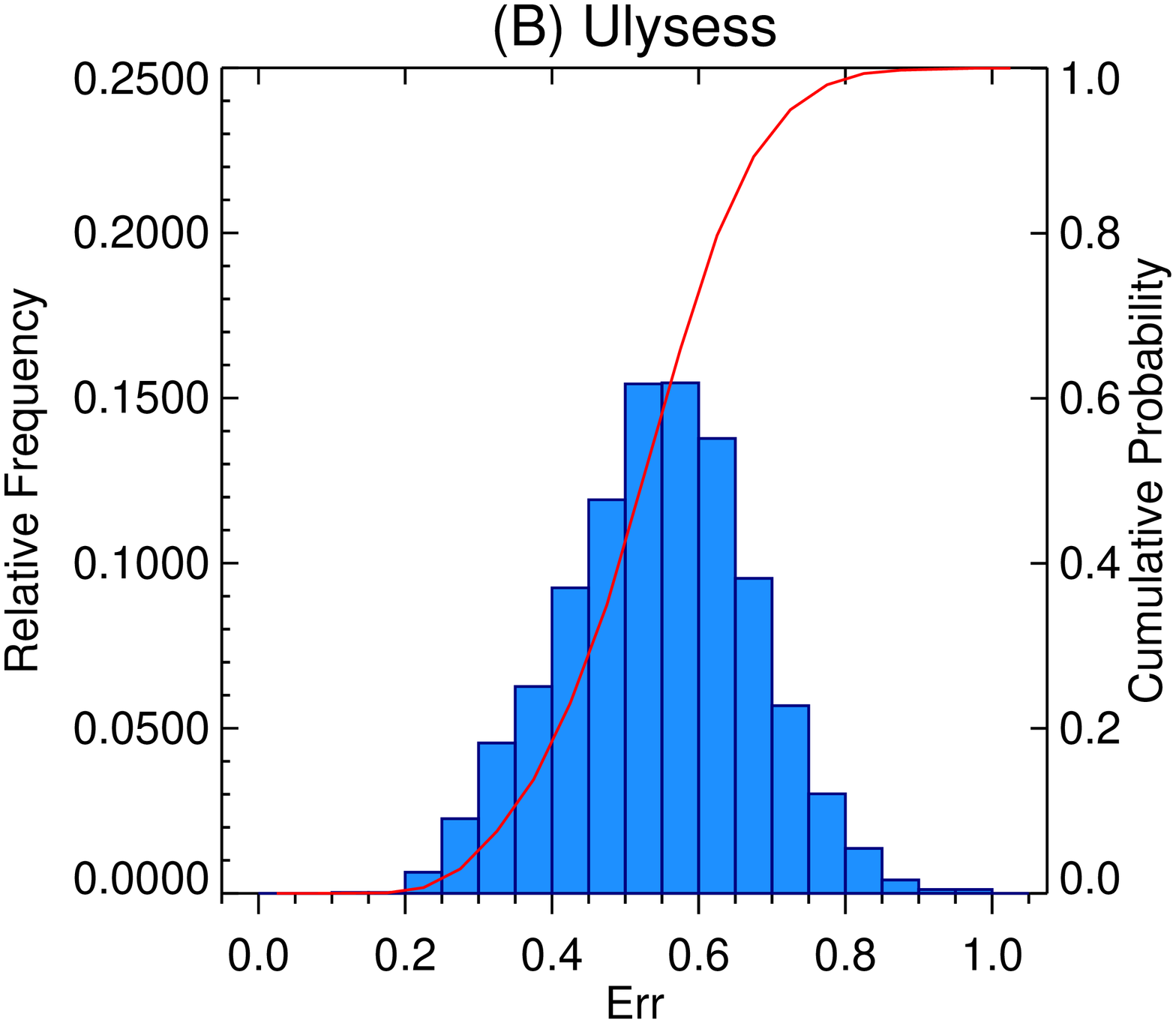}
\noindent\includegraphics[width=18.5pc,angle=0]{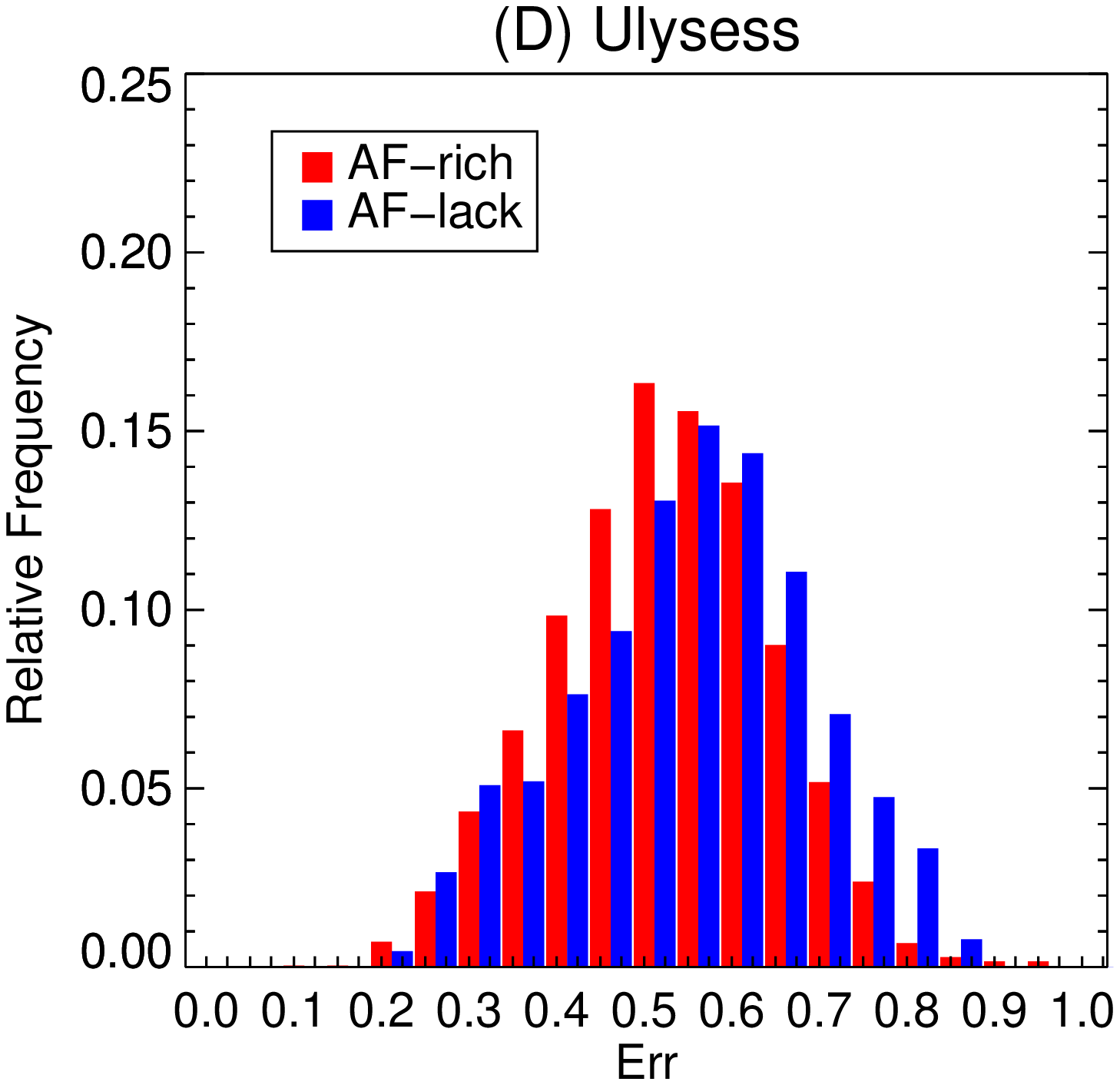}
\noindent\includegraphics[width=22pc,angle=0]{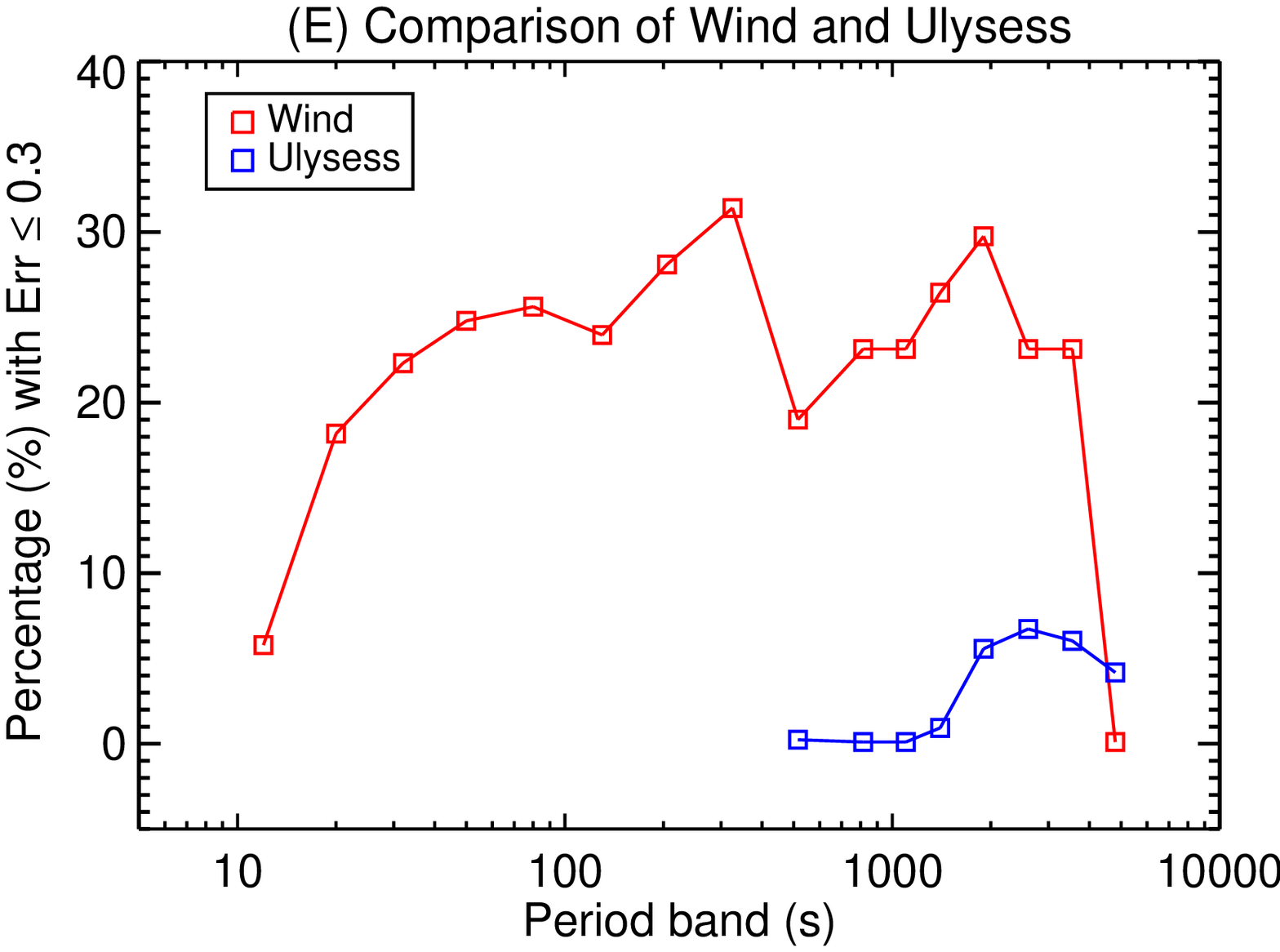}

\caption{Distribution of relative frequency of $E_{rr}$ inside the whole ICME: (A) \textit{Wind}; (B) \textit{Ulysess}. The red line represents the cumulative probability. Comparison of relative frequency distribution of $E_{rr}$ at the AF-rich region and at the AF-lack region: (C) \textit{Wind}; (D) \textit{Ulysess}. Comparison of the distribution of percentage with $E_{rr} \leq 0.3$ at different period bands are shown in panel (E).}
\label{pdf}
\end{figure}

Figure~\ref{pdf} shows the distribution of relative frequency of $E_{rr}$ inside the ICME. For the ICME observed by \textit{Wind} at 1 au, the relative frequency of $E_{rr}$ represents a bimodal distribution (panel (A)), with one peak at $\sim$ 0.3 and the other one at $\sim$ 0.6. The cumulative probability for $E_{rr} \leq 0.3$ is 21.7\%. Panel (C) shows the comparison of relative frequency distribution of $E_{rr}$ at the AF-rich region and at the AF-lack region. Different from the result for the whole ICME, the relative frequency distributions of $E_{rr}$ at both the AF-rich region and the AF-lack region have a unimodal distribution, while the peak for the AF-rich region is at $\sim 0.25$ and the peak for the AF-lack region is at $\sim 0.55$. The cumulative probability for $E_{rr} \leq 0.3$ is 36.0\% and 0.1\%, respectively. 

Panel (B) and (D) show the results inside the ICME observed by \textit{Ulysess} at 5.4 au. Different from the bimodal distribution at 1.0 au, the relative frequency of $E_{rr}$ inside the whole ICME represents a unimodal distribution with a peak at $\sim 0.55$. The cumulative probability for $E_{rr} \leq 0.3$ is only 3.0\%. Meanwhile, the detailed distributions at both the previous AF-rich region and the AF-lack region represent a similar unimodal distribution with a peak at $\sim 0.55$. The cumulative probability for $E_{rr} \leq 0.3$ is 2.9\% and 3.4\%, respectively. 

Panel (E) shows the comparison of the distribution of percentage with $E_{rr} \leq 0.3$ at different period bands. It is clear that the percentage of AFs with high degree of Alfve\'nicity at 1.0 au is much larger than that at 5.4 au. For the ICME at 1.0 au, the percentages with $E_{rr} \leq 0.3$ are more than 20\% with the period band from 30 s to 4000 s. However, those percentages are nearly 1\% with the period band from 500 to 1000 s and are less than 8\% with the period band from 2000 to 5000 s.

\section{Energetics Analysis of the ICME}
\label{energy}
The magnetic fluctuations inside an ICME represented a power spectrum in the form of $f^{-5/3}$ at spacecraft-frame frequencies less than 0.5 Hz \citep{Leamon et al 1998}. Based on the Kolmogoroff's theory, such an inertial range spectrum indicates strong spectral energy transfer. The turbulence cascade rate ($\varepsilon_{ko}$) can be deduced from the Kolmogoroff spectrum \citep{Coleman Jr 1968,Leamon et al 1999,Liu et al 2006} in the expression 
\begin{linenomath}
\begin{equation}
   \varepsilon_{ko} = \left(\frac{5}{3C_{ko}}\right)^{3/2}\frac{2\pi}{\nu}f^{5/2}[P(f)]^{3/2}
\end{equation}
\end{linenomath}
$C_{ko}$ is a numerical constant, which is assumed to be 1.6 here \citep{Batchelor 1953}. $\nu$ is the solar wind bulk speed. $P[f]$ is the observed frequency spectrum in the inertial range, and should be scaled to velocity (Alfv\'en) units \citep{Leamon et al 1999}. Here, we will follow the approach carried out by \citet{Leamon et al 1999,Liu et al 2006} to determine the turbulence dissipation rate inside the ICME.

\begin{figure}[htbp!]
\centering
\noindent\includegraphics[width=18.5pc,angle=0]{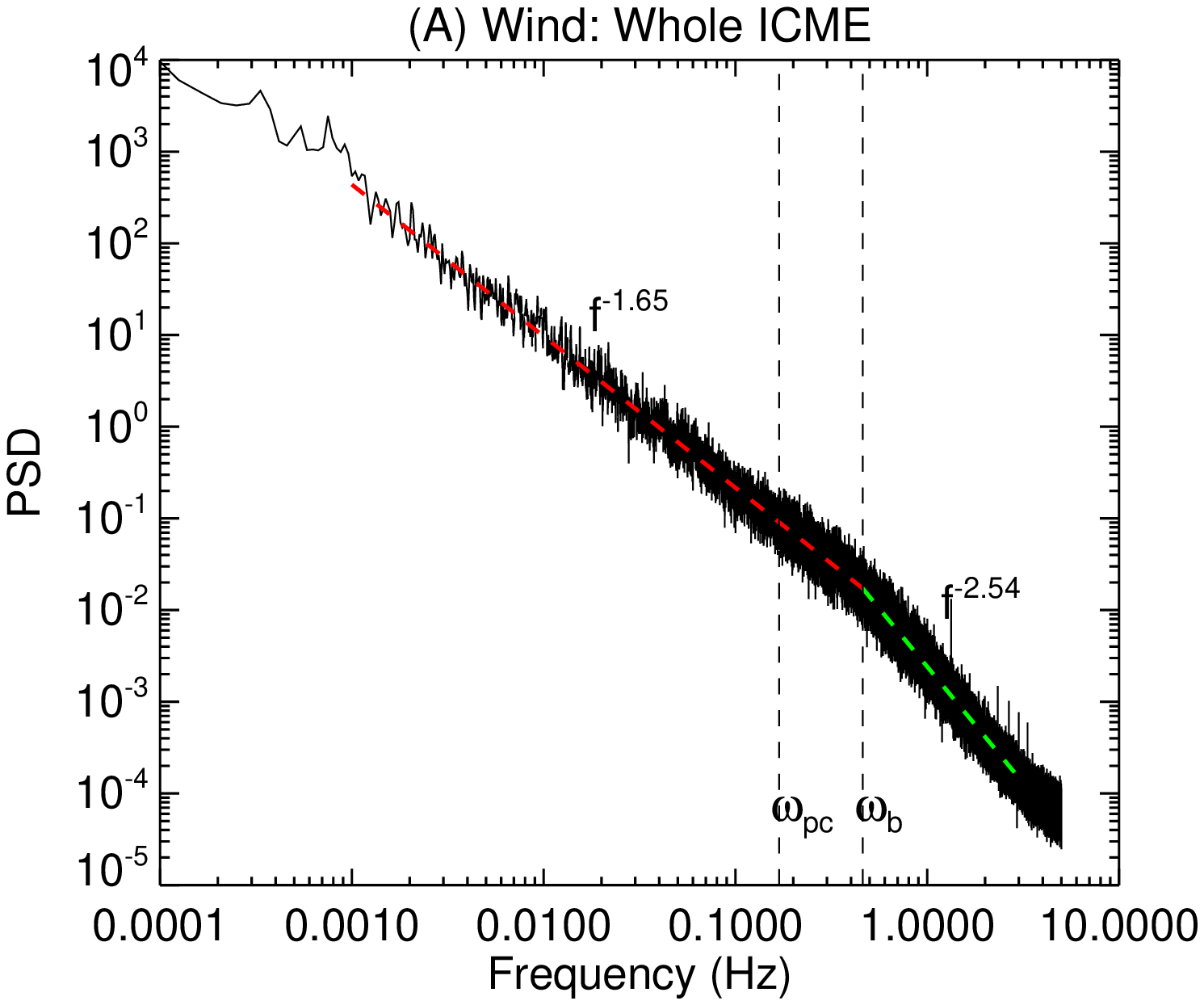}
\noindent\includegraphics[width=18pc,angle=0]{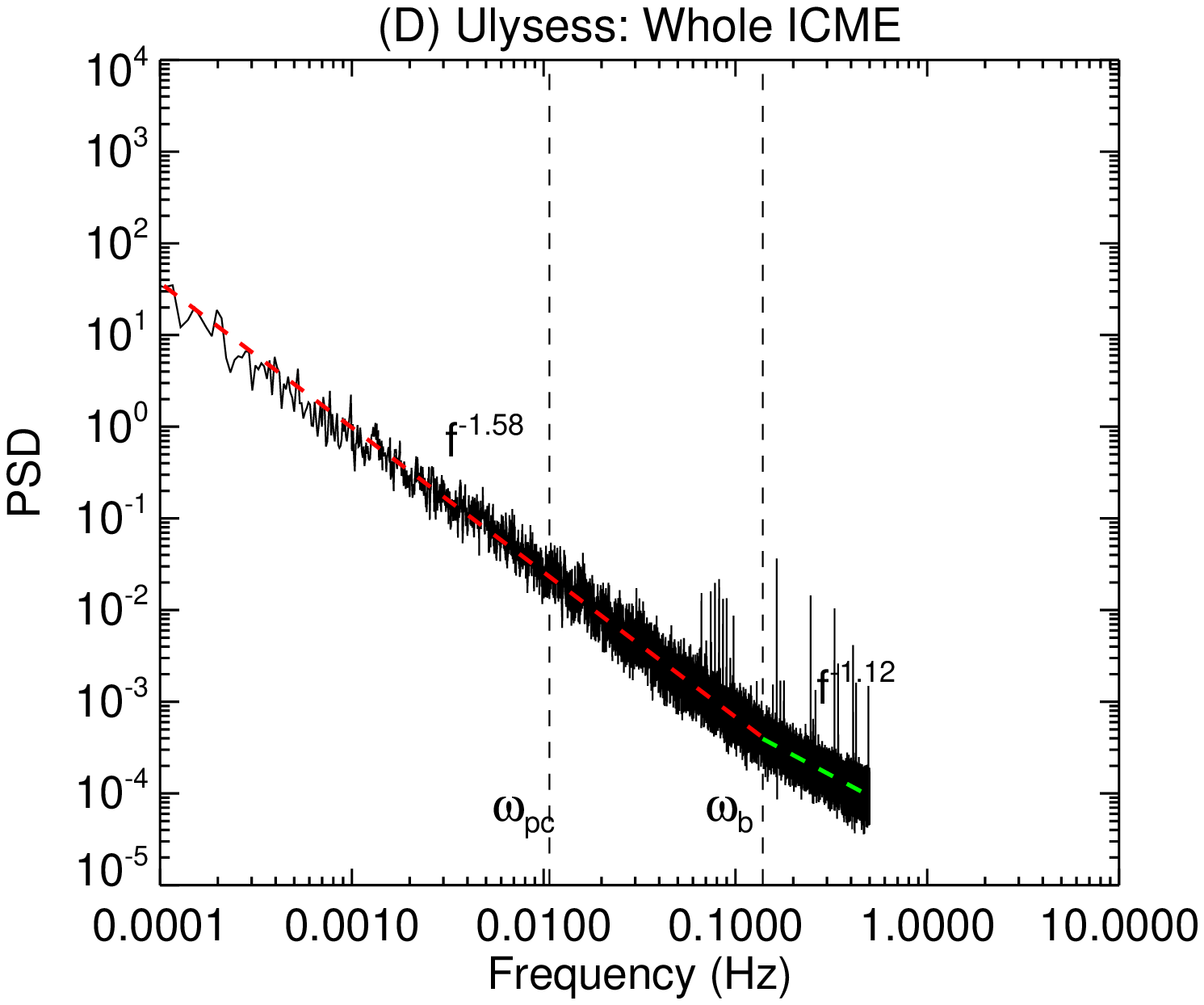}
\noindent\includegraphics[width=18.5pc,angle=0]{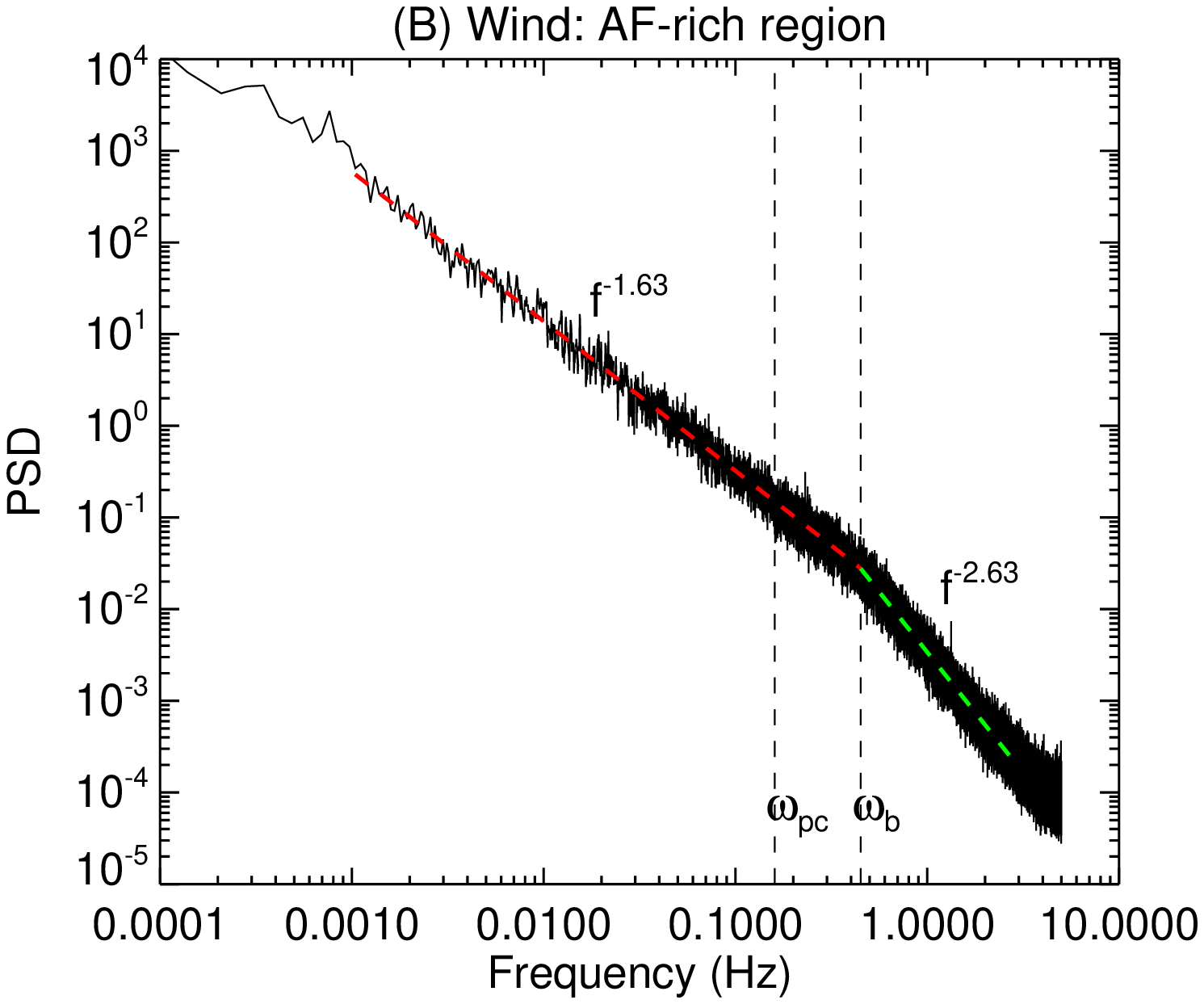}
\noindent\includegraphics[width=18pc,angle=0]{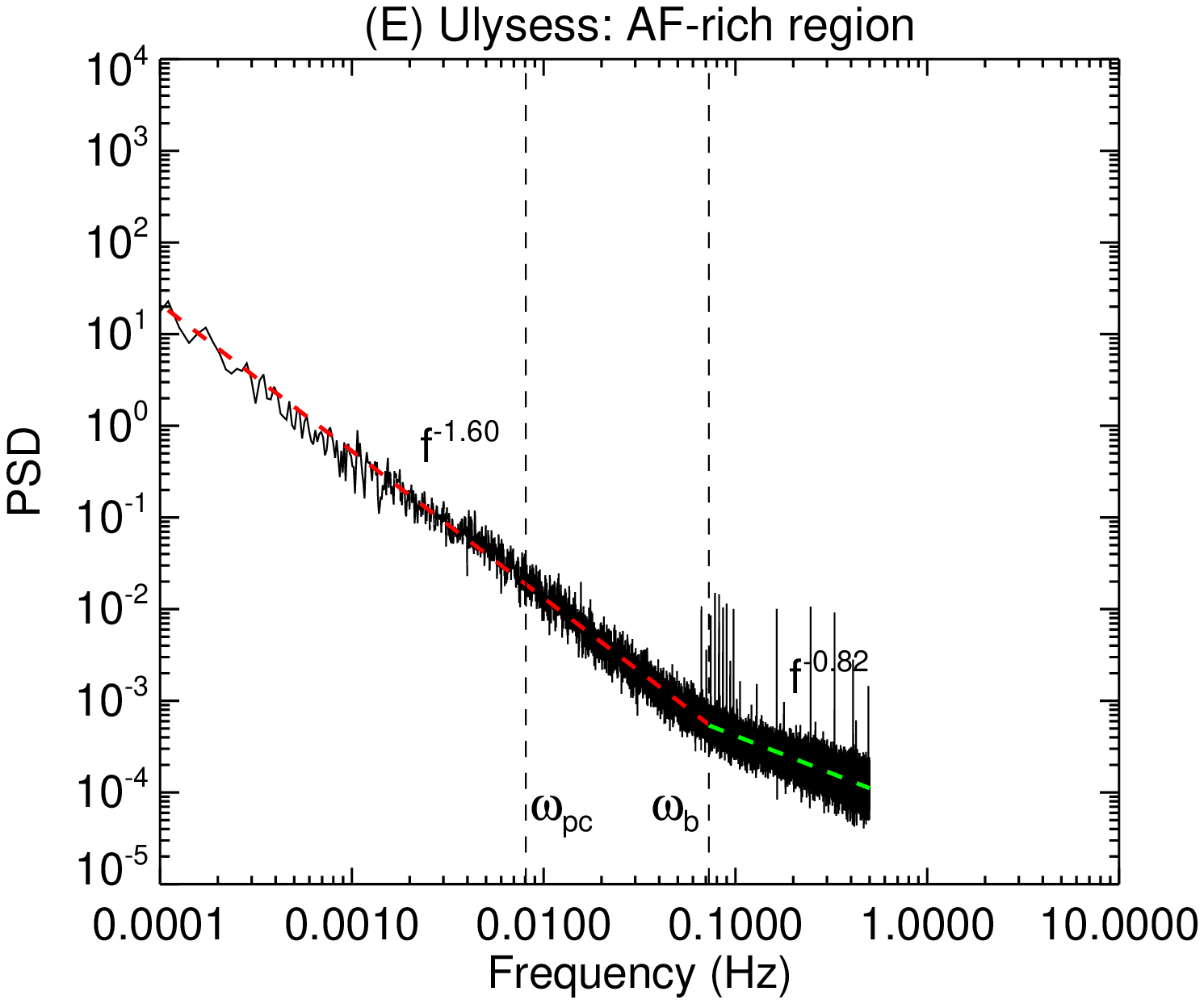}
\noindent\includegraphics[width=18.5pc,angle=0]{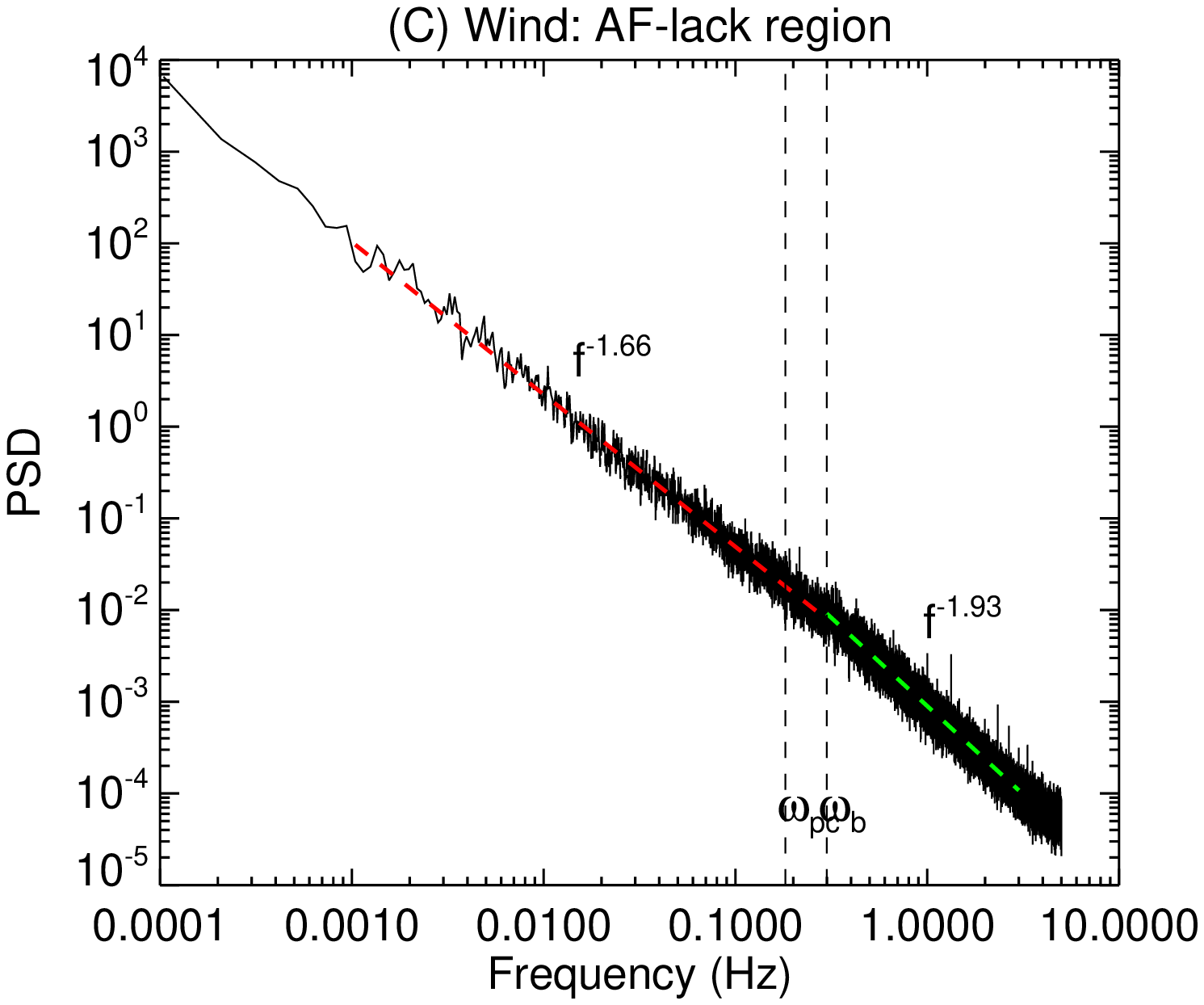}
\noindent\includegraphics[width=18pc,angle=0]{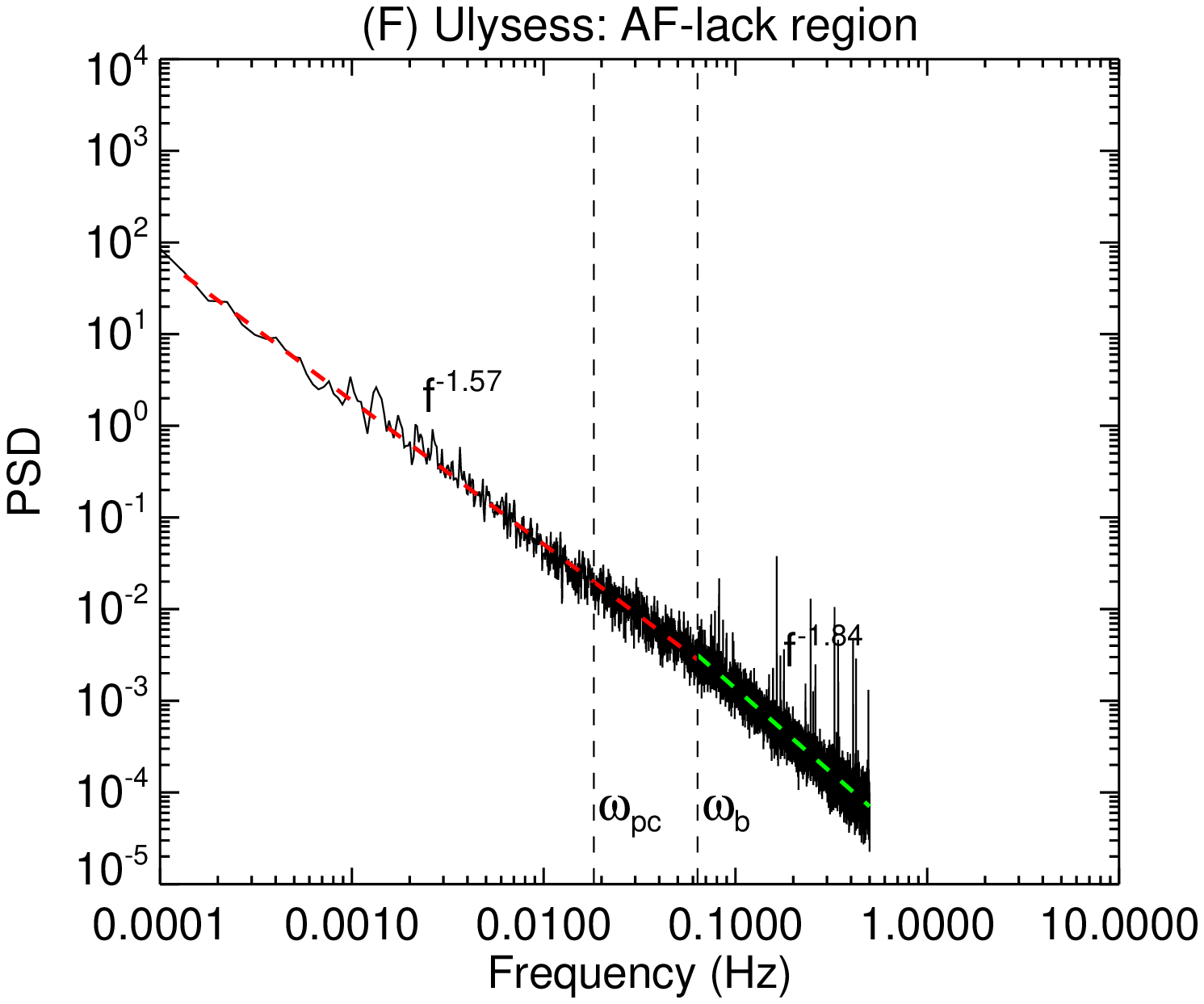}

\caption{Power spectral density of magnetic fluctuations inside the ICME: (A) the whole ICME observed by \textit{Wind} at 1.0 au; (B) the AF-rich region of the ICME at 1.0 au; (C) the AF-lack region of the ICME at 1.0 au; (D) the whole ICME observed by \textit{Ulysess} at 5.4 au; (B) the AF-rich region of the ICME at 5.4 au; (C) the AF-lack region of the ICME at 5.4 au. The dashed lines are the power law fitting results. The vertical dashed lines represent the proton gyro-frequency, $\omega_{pc}$, and the break frequency, $\omega_{b}$.}
\label{psd}
\end{figure}

Figure \ref{psd} shows the power spectral density of magnetic fluctuations inside the ICME. The left three panels are for the ICME observed by \textit{Wind} at 1.0 au. The power spectra show significant steepening at high frequencies, which marks the onset of magnetic dissipation \citep{Leamon et al 1999}. For the whole ICME, the power spectra ``break'' from a $f^{-1.65}$ power law in the inertial range to a $f^{-2.54}$ power law in the dissipation range. For the AF-rich and AF-lack regions, the spectral indexes in the inertial range are 1.63 and 1.66, respectively, in good agreement with the Kolmogoroff prediction of 5/3; meanwhile, the spectral indexes in the dissipation range are 2.63 and 1.93, respectively, indicating that the spectral cascade in the AF-rich region tends to be higher than that in the AF-lack region. The break frequency is about 0.3 $\sim$ 0.5 Hz, larger than the proton gyro-frequency, $\omega_{pc} \sim 0.2$ Hz, indicating the presence of proton cyclotron damping process. The right three panels are for the ICME observed by \textit{Ulysess} at 5.4 au. The power spectra for the whole ICME, the previous AF-rich and AF-lack region, nearly obey the Kolmogoroff theory, with the spectral indexes in the inertial range of 1.58, 1.60 and 1.57, respectively. However, the power spectra have up-wrapped for the whole ICME and the previous AF-rich region, with the spectral indexes of 1.12 and 0.82, respectively. The power spectrum for the previous AF-lack region still has a steepening in the dissipation range, with the spectral index of 1.84. Similarly, the break frequency is also larger than the proton gyro-frequency.

The energy cascade rate then can be calculated from equation (1). For the whole ICME and the AF-rich region, the energy cascade rates at 1.0 au are derived to be about 1622.3 and 2220.9 $J\cdot kg^{-1}\cdot s^{-1}$, respectively. However, when the ICME propagates to 5.4 au, those values reduce to 153.4 and 48.8 $J\cdot kg^{-1}\cdot s^{-1}$, respectively, suggesting that the capacity of turbulence cascade reduces as the ICME propagates outward due to AFs dissipation. For comparison, the energy cascade rates for the AF-lack region at 1.0 au and 5.4 au are 293.7 and 377.8 $J\cdot kg^{-1}\cdot s^{-1}$, respectively, indicating that the turbulence cascade rate can maintain at a certain level if there is lack of AFs, although the magnetic filed intensity decreases from 1.0 au to 5.4 au.

\citet{Liu et al 2006} has taken into account the Coulomb energy transfer between protons and alphas, and derived the equations of the heating rate required for protons ($\varepsilon_p$) and alphas ($\varepsilon_\alpha$) to produce the observed temperature profile. The specific expressions are as follows:
\begin{linenomath}
\begin{equation}
   \varepsilon_p = \frac{k_BT_p}{m_p}\left[\frac{3\nu}{2}\frac{d}{dr}\ln T_p-\frac{3\left(\frac{T_\alpha}{T_p}-1\right)}{2\tau_{p\alpha}}+\frac{1}{\tau_e}\right]
\end{equation}
\end{linenomath}
\begin{linenomath}
\begin{equation}
   \varepsilon_\alpha = \frac{k_BT_\alpha}{m_\alpha}\left[\frac{3\nu}{2}\frac{d}{dr}\ln T_\alpha-\frac{3\left(\frac{T_p}{T_\alpha}-1\right)}{2\tau_{\alpha p}}+\frac{1}{\tau_e}\right]
\end{equation}
\end{linenomath}
where $k_B$ is the Boltzmann constant, $T_p$ and $T_{\alpha}$ are the temperature of protons and alphas, $m_p$ and $m_{\alpha}$ are the masses of protons and alphas, $\tau_{p\alpha}$ and $\tau_{\alpha p}$ are the Coulomb collision timescales and have a relationship of $\frac{\tau_{\alpha p}}{\tau_{p\alpha}}=\frac{n_\alpha}{n_p}$, $n_\alpha$ and $n_p$ are the number densities of protons and alphas, $\tau_e$ is the expansion time of the plasma. For more details please refer to the Appendix A in \citet{Liu et al 2006}. The total required heating rate ($\varepsilon_{re}$) is then obtained as
\begin{linenomath}
\begin{equation}
   \varepsilon_{re} = \varepsilon_p+\varepsilon_\alpha
\end{equation}
\end{linenomath}

Based on the observations at 1.0 au and 5.4 au, we can quantitatively estimate the required heating rate, $\varepsilon_{re}$, to be about 1653.2 $J\cdot kg^{-1}\cdot s^{-1}$, which is in agreement with the statistical estimation made by \citet{Liu et al 2006}, 2550 $J\cdot kg^{-1}\cdot s^{-1}$. At the same time, this value is also comparable to the turbulence cascade rate inside the whole ICME at 1.0 au, 1622.3 $J\cdot kg^{-1}\cdot s^{-1}$. 

Table 1 gives a summary of some properties of the ICME observed at 1.0 au and 5.4 au. From 1.0 au to 5.4 au, the width of the ICME estimated from ICME duration and propagating speed increased from 0.26 au to 0.94 au, about 3.6 times. Accordingly, the area of the ICME cross section increases by nearly 13.1 times in the hypothesis of circular cross section. Considering the conservation of magnetic flux, the magnetic field strength should have a about 13.1 times decrease. For the AF-lack region, the magnetic field intensity changes from 11.9 nT to 1.19 nT, 10 times, which satisfies the conservation of magnetic flux. However, for the AF-rich region, the magnetic field intensity changes from 10.5 nT to 0.53 nT by 19.8 time, which is much larger than expected. This indicates the existence of an extra magnetic dissipation mechanism. At 1.0 au, the turbulence cascade rate at the AF-rich region is much larger than that at the AF-lack region because of the existence of AFs. With the disappear of AFs at the AF-rich region from 1.0 au to 5.4 au, the turbulence cascade rate decays significantly, from 2220.9 to 48.8 $J\cdot kg^{-1}\cdot s^{-1}$. However, the turbulence cascade rates at the AF-lack region can maintain a certain level, indicating such an extreme decrease is not simply caused by the decrease of the magnetic filed intensity. AFs dissipation should play a key role in the energy transfer. Based on the above results, we suggest that AFs dissipation inside ICMEs might be responsible for extra magnetic dissipation, and the turbulence cascade rate is enough to supply the required heating rate.

\begin{table}[h!]
\renewcommand{\thetable}{\arabic{table}}
\caption{A summary of some properties of the ICME observed at 1.0 au and 5.4 au.}
\begin{center}
\begin{tabular*}{36pc}{c@{\extracolsep{\fill}}c@{\extracolsep{\fill}} c@{\extracolsep{\fill}}c@{\extracolsep{\fill}}c@{\extracolsep{\fill}}}

\hline
\hline
 & \multicolumn2c{\textit{Wind}} & \multicolumn2c{\textit{Ulysess}} \\
\cline{2-3} \cline{4-5}
 & AF-rich Region & AF-lack Region & AF-rich Region & AF-lack Region \\
\hline
$Width$ (AU) & \multicolumn2c{0.26} & \multicolumn2c{0.94} \\
\hline
$|B|$ (nT) & 10.5 & 11.9 & 0.53 & 1.19 \\
\hline 
AF occurrence & 36\% & 0.1\% & 2.9\% & 3.4\% \\ 
\hline 
$\varepsilon_{ko}$ ($J\cdot kg^{-1}\cdot s^{-1}$) & 2220.9 & 293.7 & 48.8 & 377.8 \\
\hline
$\varepsilon_{re}$ ($J\cdot kg^{-1}\cdot s^{-1}$) & \multicolumn4c{1653.2}\\
 
\hline
\hline
\end{tabular*}
\end{center}
\label{comparison}
\end{table}

\section{Summary}
In this study, we track an ICME from 1.0 au to 5.4 au. Such an event could provide us a good opportunity to study the evolution of embedded AFs within an ICME and their contributions to local plasma heating directly. The ICME at 1.0 au could be divided into two regions according to the occurrence of AFs. The first one contains many AFs with high degree of Aflv\'enicity at broadband frequencies from 4$\times 10^{-4}$  to 5$\times 10^{-2}$ Hz, which is referred as the AF-rich region. The other part is lack of AFs which is thus called the AF-lack region. When the ICME propagates to 5.4 au, the Aflv\'enicity decreases significantly and the AFs at the AF-rich region are only found at few localized frequencies. The occurrence rate of AFs inside ICME at 5.4 au decreases to 3.0\% from 21.7\% at 1.0 au. Because of ICME expansion effect, the magnetic field intensity has decrease by 10.0 times for the AF-lack region. However, it decreases by almost 19.8 times for the AF-rich region, indicating the existence of an extra magnetic dissipation. 

We also estimate the energetics of the ICME at different radial distance. Under a similar magnetic field intensity situation at 1.0 au, the turbulence cascade rate estimated from the inertial range power spectrum of magnetic fluctuations within the AF-rich region is 2220.9 $J\cdot kg^{-1}\cdot s^{-1}$, which is much larger than the value of 293.7 $J\cdot kg^{-1}\cdot s^{-1}$ at the AF-lack region. If there is lack of AFs, the turbulence cascade rate can maintain a certain level, from 293.7 to 377.8 $J\cdot kg^{-1}\cdot s^{-1}$, as the decrease of magnetic field intensity from 1.0 au to 5.4 au. However, when there exists many AFs, it reduces significantly from 2220.9 to 48.8 $J\cdot kg^{-1}\cdot s^{-1}$ as the AFs disappear. The turbulence cascade dissipation rate within the whole ICME at 1.0 au is inferred to be 1622.3 $J\cdot kg^{-1}\cdot s^{-1}$, which satisfies the requirement of local ICME plasma heating rate, 1653.2 $J\cdot kg^{-1}\cdot s^{-1}$. We suggest that AF dissipation is responsible for extra magnetic dissipation and local plasma heating inside ICME.

\acknowledgments

We thank the \emph{Wind} and \emph{Ulysess} Mission for the use of their data. The data are accessible at the NSSDC (\url{ftp://nssdcftp.gsfc.nasa.gov}). We also thank Dr. J. S. He from Peking University for many valuable discussions. This work was supported by NNSFC grants 41574169 and Young Elite Scientists Sponsorship Program by CAST, 2016QNRC001. H. Li was also supported by Youth Innovation Promotion Association of the Chinese Academy of Sciences and in part by the Specialized Research Fund for State Key Laboratories of China.

\end{document}